# Self-assembled Behaviors of Desulphurized $MoS_2$ Monolayer Sheets

Pinqiang Cao,[1, 2, *] Jianyang Wu,[2, *]

[1] School of Resources and Environmental Engineering, Wuhan University of Science and Technology, Wuhan 430081, PR China

[2] Department of Physics, Research Institute for Biomimetics and Soft Matter, Jiujiang Research Institute, Fujian Provincial Key Laboratory for Soft Functional Materials Research, Xiamen University, Xiamen 361005, PR China

Email address:

Pinqiang Cao: pinqiang@wust.edu.cn;

Jianyang Wu: jianyang@xmu.edu.cn;

*To whom correspondence should be addressed: pinqiang@wust.edu.cn; jianyang@xmu.edu.cn

**Abstract:** Self-assembled topological structures of post-processed two-dimensional materials exhibit novel physical properties distinct from those of their parent materials. Herein, the critical role of desulphurization on self-assembled topological morphologies of molybdenum disulfide ($MoS_2$) monolayer sheets is explored using molecular dynamics (MD) simulations. MD results show that there are differences in atomic energetics of $MoS_2$ monolayer sheets with different desulphurization contents. Both free-standing and substrate-hosted $MoS_2$ monolayer sheets show diversity in topological structures such as flat surface, wrinkles, folds and scrolls, depending on the desulphurization contents, planar dimensions and ratios of length-to-width of $MoS_2$ monolayer sheets. Particularly, at the critical desulphurization contents, they roll up into nanotube morphology, consistent with previous experimental observations. Moreover, the observed differences in the molecular morphological diagrams between free-standing and substrate-hosted $MoS_2$ monolayer sheets can be attributed to unique interatomic interactions and van der Waals interactions in them. The study provides important insights into functionalizing structural morphological properties of two-dimensional materials, *e.g.*, $MoS_2$, via defect engineering.

## 1 Introduction

Monolayer two-dimensional (2D) transition metal dichalcogenides (TMDs) structures have attracted significant attention as promising materials due to their important practical applications in the fields of electronic devices, energy storage, biosensors, bioimaging technologies, and catalysis [1-6]. Particularly, among the TMDs' family, molybdenum disulfide ($MoS_2$) has attracted great attention because of its unique crystal structures and distinctive excellent physical properties, *e.g.*, electrical, optical, and mechanical properties.

To date, there are three most commonly encountered crystalline polymorphs of $MoS_2$ structures, namely, 1T, 2H, and 3R, where the letters stand for trigonal, hexagonal and rhombohedral structure, respectively, and the digit numbers indicate the layer number in the basic unit cell of $MoS_2$ structures [6, 7]. For each layer of the three polymorphs of $MoS_2$, it is structurally characterized by S-Mo-S sandwiched structures and each transition molybdenum (Mo) atom is covalently bonded to six sulfur (S) atoms. For bulk structures, S-Mo-S layers are connected together by weak interlayer van der Waals (vdW) interactions [8]. Importantly, molybdenum disulfide as a natural mineral is naturally abundant on Earth [9, 10], and it is predominantly available in the 2H phase with trigonal prismatic coordination of Mo atoms. Synthetic $MoS_2$ may occur in the 3R phase and 1T phase as lithium intercalation is introduced [11]. Critically, due to its unique structure, $MoS_2$ shows unique transport, electronic and mechanical properties. For example, monocrystalline $MoS_2$ monolayer exhibits a larger intrinsic direct bandgap of 1.8 eV than that of its bulk and other 2D material counterparts [9, 12, 13]. Moreover, the mobility of a room-temperature $MoS_2$ monolayer was demonstrated to have at least 200 $cm^2 \bullet V^{-1} \bullet s^{-1}$, and transistors with a current on/off ratio exceeding $1 \times 10^8$ at room-temperature can be achieved [10]. Mechanically, monolayer $MoS_2$ shows high Young's modulus and high flexibility [14-18]. Interestingly, structural transformations between different crystal structures of $MoS_2$ have been observed [19-23].

Remarkably, topological defects such as vacancies, dislocations, and grain boundaries (GBs) in $MoS_2$ monolayer sheets play an important role in tuning their transport, electronic and mechanical properties [24-27]. In reality, as a result of extra-large ratios of surface-to-bulk in monolayer $MoS_2$, topological defects are easily tended to be generated [28]. For example, sulfur vacancies, hereafter also called as desulphurization in this work, in the basal plane of $MoS_2$ is one common type of topological defects. Such desulphurization induced topological defects can be introduced by ion sputtering and sulfur annealing [25, 29, 30]. The introduction of sulfur vacancies can enhance hydrogen evolution reaction activity of monolayer $MoS_2$ [25, 31-34]. In terms of magnetics [35], anion and cation vacancies and vacancy clusters in $MoS_2$ can result in ferromagnetic effects. In view of these interesting phenomena, it suggests that material performances of molybdenum disulfide can be modulated via topological defect engineering.

As a result of its large out-of-plane flexibility, 2D layered $MoS_2$ shows different surface morphologies at various external conditions [29, 36]. For example, when a pre-stretched strain in the elastomeric substrate is suddenly released, well-aligned wrinkles of layered $MoS_2$ sheets can form as a consequence of buckling-induced delamination [37, 38]. During mechanical exfoliation of $MoS_2$ sheets, few layer $MoS_2$ flakes spontaneously form wrinkles and are buckling-delaminated on elastomer substrates due to large elastic mismatch. Monolayer $MoS_2$ can form wrinkles under uniaxial mechanical loadings [39], which is also observed in graphene, h-boron nitride (h-BN), and black phosphorus (BP) [18, 40-42]. Individual wrinkles in few-layer $MoS_2$ sheets were also observed as they are lithiated or mechanically processed, and these individual wrinkles are analogous to line dislocations in bulk materials, but they showed drastically different interaction behaviors. Intrinsic ripples in other 2D materials like graphene spontaneously appear due to thermal fluctuations [31, 35]. Also, it was identified that S vacancies can lead to the formations of wrinkles of 2H $MoS_2$ [29]. Despite a number of important findings focusing on the surface morphology of $MoS_2$ sheets have been reported, yet how the desulphurization affects the structural morphology of $MoS_2$ sheets remains unclear at molecular scale, particularly for the structural morphology phase versus desulphurization contents. In this work, a series of molecular dynamics (MD) simulations are systematically performed to reveal the critical role of desulphurization on the structural morphology of monocrystalline $MoS_2$ monolayer sheets.

## 2 Methods

### 2.1 Molecular Models

Monocrystalline $MoS_2$ monolayer samples are generated to investigate desulphurization-induced changes in structural morphology of $MoS_2$. Unlike other 2D materials, *e.g.*, flat graphene, puckered honeycomb BP, $MoS_2$ possesses a S-Mo-S sandwiched structure. The unit cell of $MoS_2$ is composed of one molybdenum atom plus six sulfur atoms. Six S atoms form triangular prism with a Mo atom located in the center, then Mo and S atoms covalently share electrons. Contrarily, the different layers of S-Mo-S are weakly connected to each other through vdW interactions in the bulk $MoS_2$. The thickness of monolayer $MoS_2$ is approximately 0.65 nm [10]. Figures 1a and 1b show different viewed snapshots of monolayer and bilayer $MoS_2$ structures, where Mo and S atoms are red- and yellow-colored, respectively. Figures 1c and 1d show perspective snapshots of free-standing $MoS_2$ monolayer sheets and substrate-hosted $MoS_2$ monolayer sheets on large-area $MoS_2$ monolayer sheets with an in-plane dimension of $10 \times 10$ nm$^2$, respectively. Figure 1e shows a perspective snapshot of zoomed-in $MoS_2$ bilayers. To investigate the effect of S vacancies on the structural morphology of $MoS_2$ layers, desulphurization contents from 0.0-100% on one planar surface of $MoS_2$ samples with dimensions varying from $10 \times 10 – 50 \times 50$ nm$^2$ and length-to-width ratios from 1:1 to 5:1 are considered for both free-standing and substrate-hosted $MoS_2$ monolayer sheets, respectively. All S vacancies (desulphurization) are generated on one surface of $MoS_2$ monolayer sheets by randomly deleting the S atoms on the one surface of $MoS_2$ monolayer. For clarity, the one surface of $MoS_2$ containing S vacancies is hereafter also called as top-surface of $MoS_2$ monolayer, and the other one surface of $MoS_2$ without containing S vacancies is hereafter also called as bottom-surface of $MoS_2$ monolayer.

### 2.2 Forcefields of $MoS_2$ Structures

A many-body reactive empirical bond order (REBO)-type Mo-S potential developed by Liang *et al.*[43, 44] is employed to describe the interatomic interactions in $MoS_2$ systems. The forcefield contains a many-body REBO potential [45] that describes the short-range interactions and a two-body Lennard-Jones potential for the long-range vdW interactions. It is capable of capturing Mo-S covalent bond breaking and reforming during molecular dynamics calculations. Furthermore, this potential has successfully been utilized to investigate physical and chemical properties of $MoS_2$ nanostructures [23, 46-53]. Importantly, it also shows good description in the structure behaviors of $MoS_2$ and dealing with the point defects in $MoS_2$ samples [52, 54]. The short-ranged interatomic interactions of Mo-S systems by REBO-type Mo-S forcefield are expressed as:

$$E_b = \frac{1}{2}\sum_i \sum_{j(>i)} \left[ V^R(r_{ij}) - b_{ij} V^A(r_{ij}) \right] \tag{1}$$

$$V^R(r) = f^C(r)(1 + Q/r)Ae^{-\alpha r} \tag{2}$$

$$V^A(r) = f^C(r)\sum_{n=1}^{3} B_n e^{-\beta r} \tag{3}$$

where $E_b$ is the total binding energy. $V^R(r_{ij})$ and $V^A(r_{ij})$ are a repulsive and an attractive reaction, respectively. $r_{ij}$ is the distance between atoms $i$ and $j$. The cutoff function $f^C(r)$ is obtained from the switching cutoff scheme. The values of parameters used in our MD simulations can be found in elsewhere [43, 44]. The long-ranged vdW interactions in $MoS_2$ system are described by the 12-6 Lennard-Jones potential as:

$$E_{LJ}(r) = 4\varepsilon\left[\left(\frac{\sigma}{r}\right)^{12} - \left(\frac{\sigma}{r}\right)^{6}\right] \quad (4)$$

where $r$ is the interatomic distance. $\sigma$ and $\varepsilon$ is chosen as 3.13Å and 0.00693 eV in this study, respectively.

### 2.3 MD Simulations

All the MD calculations are carried out using the Large-scale Atomic-Molecular Massively Parallel Simulator (LAMMPS) software package [55]. At the beginning, all samples are quasi-statically relaxed to a configuration with local minimum energy through conjugate gradient method, with an energy tolerance of $1.0 \times 0^{-4}$ eV and a force tolerance of $1.0 \times 10^{-4}$ eV/Å, respectively. Then, MD relaxations are conducted with $1.0 \times 10^{6}$ timesteps at 298.15 K under the NVT (constant number of particles, constant volume, and constant temperature) ensemble. The Nosé-Hoover thermostat method with damping times of 0.1 ps is assigned to control the temperature. The velocity-Verlet integration algorithm is utilized to integrate the equations with a timestep of 1 fs in all MD simulations. Periodic boundary conditions (PBCs) are applied in all three orthogonal directions. Initial velocities of all Mo and S atoms in the $MoS_2$ samples are assigned following the Gaussian distribution at the given temperature. The thickness of monolayer $MoS_2$ is taken as 0.65 nm to express atomic stress [10]. Based on the virial definition of stress, the atomic stress is computed using the forces on the atoms collected during the MD calculations. Both potential energy and atomic stress in our samples are averaged over 1000 timesteps to eliminate the fast fluctuations.

## 3 Results and Discussion

### 3.1 Energetics in Desulphurized $MoS_2$ Sheets

Figures 2a-2c show variations in the instantaneous atomic potential energy per atom of free-standing desulphurized $MoS_2$ monolayer sheets of $10 \times 10$ nm$^2$ with MD relaxation time varying from 0.0-1.0 ns. It should be noted that random sulfur defects occur on the top-surface of $MoS_2$ monolayer sheets. As a result of S-Mo-S sandwiched structures, surface S atoms including bottom- and top-S atoms in $MoS_2$ monolayer sheets with different desulphurization contents show higher potential energy than interior Mo atoms. As the desulphurization content on the top-surface of $MoS_2$ monolayer sheets varies from 0.0-100%, there are significant changes in the instantaneous atomic potential energy per atom. As shown in Figure 2a, the instantaneous atomic potential energy per Mo atom rapidly converges to constant values with increasing relaxation time, indicating the stability of interior Mo atoms. Interestingly, the converged values of Mo atoms are monotonically increased with the increasing desulphurization content, suggesting that the stability of interior Mo atoms can be adjusted by desulphurization contents. As for both top- and bottom-S atoms (Figures 2b and 2c), however, it is identified that complex changes in the instantaneous atomic potential energy per S atom can occur. For example, there are fluctuations in the instantaneous atomic potential energy per S atom with the increasing relaxation time, although the instantaneous atomic potential energy per S atom tends to reach a constant value, suggesting complex changes in the structural morphology of $MoS_2$ monolayer sheets. Interestingly, when the desulphurization content on the top-surface of $MoS_2$ monolayer sheets changes from 0.0-100%, free-standing $MoS_2$ monolayer sheets show a transition behavior in the instantaneous potential energy per bottom-S atom (Figure 2b): Below desulphurization of 50%, the instantaneous atomic potential energy per bottom-S atom is increased with increasing desulphurization contents, whereas above which, it shows a reduction trend with increasing desulphurization contents. With regard to the case of top surface of free-standing desulphurized $MoS_2$ monolayer sheets, there is negligible difference in the instantaneous atomic potential energy per top-S atom for $MoS_2$ monolayer sheets with desulphurization of 0.0-30%, whereas above 40% desulphurization, the instantaneous atomic potential energy per top-S atoms is decreased with the increasing desulphurization. It should be noted that free-standing desulphurized $MoS_2$ monolayer sheets with planar dimensions varying from 20 × 20 to 50 × 50 $nm^2$ and length-to-width ratios from 2:1 to 5:1 show similar variations in the instantaneous atomic potential energy per atom with changing desulphurization contents (Figures S2-S4, and S7-S9). The instantaneous radius of gyration ($R_g$) is an appropriate way to quantitatively measure the structural compactness and morphology. Figure 2d presents the variations in the instantaneous radius of gyration ($R_g$) with MD relaxation simulation time. As is seen, the $R_g$ of desulphurized $MoS_2$ monolayer quickly converges to averaged values, and it varies from around 31-43 Å depending on the desulphurization contents. The free-standing $MoS_2$ monolayer sheets with less desulphurization show large values of $R_g$. Such changes in the $R_g$ of free-standing desulphurized $MoS_2$ monolayer sheets indicate their distinct structural morphology. Particularly, the perfect sawtooth-like fluctuations in the $R_g$ of $MoS_2$ sheets at the desulphurization content of 50% suggest the cyclic changes of structural morphology. Similar changes in the instantaneous radius of gyration with changing desulphurization contents are also observed in the other free-standing desulphurized $MoS_2$ monolayer sheets with dimensions varying from 20 × 20 to 50 × 50 $nm^2$ and length-to-width ratios from 2:1 to 5:1 (Figures S5 and S10).

Figures 3a-3c show the instantaneous atomic potential energy per atom of finite substrate-hosted desulphurized $MoS_2$ sheets of 10 × 10 $nm^2$ on large-area $MoS_2$ monolayer substrate as a function of MD simulation time. Apparently, there are differences in developments of the instantaneous atomic potential energy per atom of finite substrate-hosted desulphurized $MoS_2$ monolayer sheets under different desulphurization contents. Upon interior Mo atoms, the instantaneous atomic potential energy per molybdenum atom in finite substrate-hosted desulphurized $MoS_2$ monolayer sheets is initially increased with increasing relaxation time (Figure 3a). As the relaxation time reaches a critical value, however, it is reduced except for substrate-hosted desulphurized $MoS_2$ monolayer sheets with both desulphurization of 40% and 100%. With further relaxations, substrate-hosted $MoS_2$ monolayer sheets with desulphurization contents varying from 0-30% show constant values of the instantaneous atomic potential energy per Mo atom, indicating achievement of stable configurations. For the case of other desulphurization contents ranging from 40-100%, the instantaneous atomic potential energy per Mo atom is either continuously increased or decreased, relying on the desulphurization contents, indicative of metastable configurations.

As shown in Figure 3b and 3c, finite substrate-hosted desulphurized $MoS_2$ monolayer sheets with desulphurization of 0-30% shows convergence of instantaneous potential energy per surface top- or bottom-S atom, again suggesting their stable structural configurations, whereas that with desulphurization of 40-100% show monotonic decreases in the instantaneous atomic potential energy per surface top- or bottom-S atom, also indicating their metastable structures. The developments of the instantaneous atomic potential energy per atom of substrate-hosted $MoS_2$ monolayer sheets differ from that of free-standing ones, indicating that the critical effects of large-area $MoS_2$ monolayer substrates on the structural revolutions of $MoS_2$ monolayer sheets due to the weak interlayer van der Waals interactions between them. Figure 3d shows the values of $R_g$ of substrate-hosted $MoS_2$ monolayer sheets as a function of MD relaxation time. Apparently, substrate-hosted $MoS_2$ monolayer sheets with high desulphurization contents show distinct developments and $R_g$ values, which are different from those of free-standing $MoS_2$ sheets. These indicate that their different structural morphologies. These desulphurization-induced features of development of energetics and $R_g$ can be similarly observed in other substrate-hosted $MoS_2$ sheets with planar dimensions of $20 \times 20$-$50 \times 50$ nm$^2$ and length-to-width ratio of 2:1 to 5:1 (Figures S12-S15, and S17-S20).

As shown in Figures S1, S6, S11, and S16, the total energy of free-standing and substrate-hosted desulphurized $MoS_2$ monolayer sheets with different planar dimensions and ratios of length-to-width can converge to constant values, indicating that the relatively relaxed stable states of $MoS_2$ samples compared to their initial states. To understand the relatively relaxed stable states of $MoS_2$ samples, Figures 4 and 5 collect the average atomic potential energy per atom of both free-standing and substrate-hosted desulphurized $MoS_2$ monolayer sheets with different dimensions and ratios of length-to-width, respectively. The average atomic potential energy per atom in Figures 4 and 5 can be estimated from the instantaneous atomic potential energy during the MD relaxation time. As shown in Figure 4a, when the sample size changes from $10 \times 10$ - $50 \times 50$ nm$^2$, free-standing $MoS_2$ sheets with desulphurization contents of 0-60% show slight reduction tendency in the average potential energy per Mo atom except for the desulphurization of 30%, whereas for that with desulphurization contents of 60-100%, the average potential energy per Mo atom is either slightly increased or nearly kept at an average constant value. Such transition behavior in the potential energy per Mo atom suggests that free-standing $MoS_2$ sheets with different dimensions shows different structural morphologies at critical desulphurization contents. With regard to free-standing $MoS_2$ sheets with different ratios of length-to-width, it is identified from Figure 4d that the average potential energy per Mo atom is monotonically slightly reduced with the increasing ratio of length-to-width, particularly for that with desulphurization content of 100%. For the case of bottom-S atom (Figures 4b and 4e), free-standing $MoS_2$ sheets with low desulphurization contents show reduction in the average potential energy per bottom-S atom with increasing dimension sizes and ratios of length-to-width. However, those with high desulphurization content show complex changes in average potential energy per bottom-S atom with increasing dimension sizes and ratios of length-to-width. Upon the case of top-S atom (Figures 4c and 4f), free-standing $MoS_2$ sheets show either reduction tendency or other complexed behaviors in the average potential energy per top-S atom with increasing dimension sizes and ratios of length-to-width, depending on the desulphurization contents. As seen in Figure 5a and d, all substrate-hosted desulphurized $MoS_2$ monolayer sheets with different dimensions and ratios of length-to-width show a reduction trend in the average atomic potential energy per Mo atom with increasing both sample size and ratios of length-to-width, and the reduction becomes less pronounced with increasing sample sizes and ratios of length-to-width. With regard to the case of surface bottom- and top-S atoms, however, there is a critical desulphurization content in the tendency change of the average atomic potential energy per bottom- or top-S atom. For example, below the desulphurization content of 30%, all $MoS_2$ samples show negligible reductions or increments in the average potential energy per bottom-S atom, whereas above which, they show obvious rising behaviors in the average potential energy per bottom-S atom, and then the rising behaviors become less significant with the increasements of sample dimensions and ratios of length-to-width (Figure 5b and e). For the case of top-S atoms, it is found that, above desulphurization content of 30%, the average potential energy per top-S atom is monotonically increased with increasing sample dimensions and ratios of length-to-width, whereas below which, substrate-hosted $MoS_2$ samples show very slight changes in the average atomic potential energy per top-S atom. By comparison with free-standing $MoS_2$ sheets with substrate-hosted $MoS_2$ samples, for a given desulphurized $MoS_2$ sheet with a given dimension and a ratio of length-to-width, the average atomic potential energy per Mo- or S atom in substrate-hosted $MoS_2$ samples are always less than those in free-standing $MoS_2$ samples except for some cases with low desulphurization contents. This is mainly due to the fact that there are different interatomic interactions and vdW interactions between free-standing and substrate-hosted $MoS_2$ systems.

### 3.2 Desulphurization-induced Changes in Morphology of $MoS_2$ Sheets

Distinct energetics and radii of gyration in both free-standing and substrate-hosted $MoS_2$ monolayer sheets with different desulphurization contents suggest that their structural morphologies can be tuned by desulphurization. To provide more insights into desulphurization-induced change in morphology of monocrystalline $MoS_2$ monolayer sheets, the developments of structural morphology of $MoS_2$ samples are recorded during the structural relaxations. Figure 6 displays the snapshots of free-standing $MoS_2$ monolayer sheets with a planar dimension of $20 \times 20$ nm$^2$ at desulphurization contents of 10%, 20%, 30%, 40%, 70% and 90% after relaxation times of 0 and 2 ns. To highlight desulphurization-induced wrinkling in free-standing $MoS_2$ sheets, atoms are rendered on the basis of relative displacements along out-of-plane ($z$) direction. Clearly, the structural morphology of free-standing $MoS_2$ sheets can be strongly affected by desulphurization. After structural optimization (zero MD relaxation time), there are large out-of-plane atomic displacements in the edges of free-standing $MoS_2$ monolayer sheets. After MD relaxation of 2 ns, however, their planar morphologies are remarkably changed. When the desulphurization content is below 30%, free-standing $MoS_2$ sheets wrinkles with low amplitude of out-of-plane fluctuations, and the out-of-plane displacement becomes more pronounced with increasing desulphurization contents, as shown in

Figures 6a-6f. As is known, intrinsic and dynamic wrinkling of $MoS_2$ change their optoelectronic and electromechanical properties, for example, coupled electronic and optical responses of $MoS_2$ can be altered by dynamic control of wrinkling degrees of freedom [56, 57]. At desulphurization contents of 40-60%, free-standing $MoS_2$ monolayer sheets roll into a nanotube. Such interesting phenomenon was also detected from the previous experimental results [29]. When the desulphurization content is about 70%, however, the edges in free-standing $MoS_2$ monolayer sheets show high ability of wrinkling, resulting in a hierarchy triangle structure with nanotube edges, as illustrated in Figure 6j. Once the desulphurization content exceeds 80%, a unique structure of diagonal-folding can form (Figure 6l).

Figure 7 shows the snapshots of substrate-hosted $MoS_2$ monolayer sheets with a planar dimension of $20 \times 20$ nm$^2$ at desulphurization contents of 0%, 40%, 50%, 60%, 80% and 90% after relaxation times of 0 and 2 ns, where atoms are colored based on the relative atomic displacements along the out-of-plane ($z$) direction. Similarly, the desulphurization content plays a critical role in the structural morphology of substrate-hosted $MoS_2$ monolayer sheets. At low desulphurization contents of 0-20%, substrate-hosted $MoS_2$ monolayer sheets well stick to the substrate of $MoS_2$ monolayer as a result of vdW interactions between the substrate-hosted $MoS_2$ monolayer and the large-area $MoS_2$ monolayer substrate, resulting in a flat morphology of substrate-hosted $MoS_2$ monolayer sheets. As the desulphurization content reaches about 30-40%, quadrangular corners in substrate-hosted $MoS_2$ monolayer sheets are folded, forming localized wrinkling structures; whereas the interior bulk $MoS_2$ monolayer still sticks to the large-area $MoS_2$ monolayer substrate, remaining a flat morphology. Once the desulphurization content reaches around 50%, however, the substrate-hosted $MoS_2$ monolayer sheets are not able to remain flat morphology, but roll into structural morphology of a nanotube, similar to the case of free-standing $MoS_2$ sheets with a planar dimension of $20 \times 20$ nm$^2$ at desulphurization contents of 40-60%. When the desulphurization content is 60%, the substrate-hosted $MoS_2$ monolayer sheets form a triangle wrinkled nanotube structure, analogous to the case of free-standing $MoS_2$ sheet with a planar dimension of $20 \times 20$ nm$^2$ at the desulphurization content of 70%. As the desulphurization content ranges from 70-80%, the substrate-hosted $MoS_2$ monolayer sheets roll into a closed diagonal-folding nanotube structure that shows non-uniform diameter along the wrinkling axis. Once the desulphurization content reaches 90%, the edges of substrate-hosted $MoS_2$ monolayer sheets promptly roll into small wrinkled nanotube structures, forming wrinkled quadrangle with nanotube edges. Obviously, the substrate-hosted $MoS_2$ monolayer sheets show more diverse structural morphologies via desulphurization than those of free-standing $MoS_2$ monolayer sheets.

### 3.3 Desulphurization-controlled Morphological Diagrams

Figure 8 summaries the morphological diagrams of free-standing desulphurized $MoS_2$ monolayer sheets with different planar dimensions and ratios of length-to-width, and here four distinct structural morphologies are defined as illustrated by the insets. As is seen from Figure 8a, the free-standing $MoS_2$ monolayer sheets with dimensions from $10 \times 10$ to $50 \times 50$ nm$^2$ show wrinkle morphology at low desulphurization contents, nanotube morphology at intermediate desulphurization contents, and diagonal-folding morphology at high desulphurization contents. Interestingly, a novel triangular-tube morphology can be only detected in free-standing $MoS_2$ monolayer sheets with dimensions of $20 \times 20$ - $40 \times 40$ nm$^2$ at intermediate desulphurization contents. The critical desulphurization content of changing structural morphology of free-standing $MoS_2$ monolayer sheets from one to another depends on the dimension sizes. As is shown in Figure 8b, all free-standing $MoS_2$ monolayer sheets with ratios of length-to-width from 2:1 to 5:1 primarily form wrinkle and nanotube morphology, which are also observed in $MoS_2$ sheets with dimensions of $10 \times 10$ - $50 \times 50$ nm$^2$. Moreover, only free-standing $MoS_2$ monolayer sheets with a length-to-width ratio of 1:1 exhibits the diagonal-folding structural morphology at high desulphurization contents. However, free-standing $MoS_2$ monolayer sheets with ratios of length-to-width of 2:1 - 5:1 are not able to show triangular-nanotube morphology through desulphurization, differing from that the case of square-shaped samples.

Figure 9 summaries the morphological diagrams of substrate-hosted desulphurized $MoS_2$ monolayer sheets with different planar dimensions and ratios of length-to-width, and here seven distinct structural morphologies are illustrated by the insets for clarity. As displayed in Figure 9a, substrate-hosted $MoS_2$ monolayer sheets with planar dimensions of $10 \times 10$ - $50 \times 50$ nm$^2$ is able to show flat morphology at low desulphurization contents. With the increasing desulphurization contents, substrate-hosted $MoS_2$ monolayer sheets can form wrinkles, nanotube, diagonal-folding, or triangular-tube morphologies depending on the planar dimensions of substrate-hosted $MoS_2$ monolayer sheets. At high desulphurization contents, substrate-hosted $MoS_2$ monolayer sheets with a planar dimension of $10 \times 10$ nm$^2$ is structurally dissociated to form fracture amorphous cluster as the desulphurization content ranges from 70-100%. In contrast, substrate-hosted $MoS_2$ monolayer sheets with dimensions of $20 \times 20$ to $50 \times 50$ nm$^2$ is able to form quadrangle-tube morphology at high desulphurization contents. These diverse morphologies, *e.g.*, wrinkles, can be controlled by desulphurization contents. Previous studies showed that the desulphurization in $MoS_2$ sheets can also introduce in-plane strains to widen the wrinkles [24, 26, 58]. It is noted that only substrate-hosted $MoS_2$ monolayer sheets with the dimension of $20 \times 20$ nm$^2$ is able to show nanotube morphology as the desulphurization content is 50%. As shown in Figure 9b, when the ratio of length-to-width is changed, substrate-hosted $MoS_2$ monolayer sheets show complex morphology diagrams similar to those of square-shaped $MoS_2$ samples. $MoS_2$ Samples with each ratio of length-to-width are able to show different distinct structural morphologies as the desulphurization content is increased from 0-100%. For example, substrate-hosted $MoS_2$ monolayer sheets with ratios of length-to-width of 2:1 - 5:1 show flat morphology at low desulphurization contents, but nanotube morphology at high desulphurization contents, differing from the case of square-shaped substrate-hosted $MoS_2$ monolayer sheets. It should be noted that substrate-hosted $MoS_2$ monolayer sheets with a length-to-width ratio of 1:1 show fracture amorphous nanocluster at high desulphurization contents.

The different structural morphology phases of desulphurized $MoS_2$ sheets as shown in Figures 8-9 indicate that the critical role of the weak interlayer van der Waals interactions on the structural morphology of desulphurized $MoS_2$. Critically, self-assembled structural behaviors of desulphurized $MoS_2$ monolayer sheets can be determined by S vacancies. On the one hand, followed by occurrence of the unique self-assembled behaviors, the stability and physical properties of $MoS_2$ can be adjusted with self-assembled new changed structural morphologies[59, 60]. More specifically, the stability of $MoS_2$ is always on the dynamic changes, which is indicated by the dynamic development of energetics and radius of gyration of desulphurized $MoS_2$ monolayer sheets. On the other hand, the introduction of S vacancies is able to result in occurrence of planar strains, substrates, *e.g.*, large-area $MoS_2$ monolayer sheets, may serve as excellent templates to accommodate the planar strains in $MoS_2$ monolayer. Interestingly, as desulphurization contents is above the critical contents, the planar strains in $MoS_2$ monolayer sheets can not be accommodated by the substrates. Followed by this case, various complex structural morphologies of desulphurized $MoS_2$ monolayer sheets appear. In a nutshell, structural morphologies of free-standing and substrate-hosted $MoS_2$ monolayer sheets are strongly dependent on the desulphurization contents, planar dimensions and ratios of length-to-width. These suggest that structural morphologies of $MoS_2$ monolayer sheets with a given planar dimension can be finely tuned via desulphurization defect engineering.

## 4 Conclusions

In summary, classical MD simulations with a many-body reactive empirical bond order potential are performed to investigate the structural morphology of desulphurized $MoS_2$ monolayer sheets. Both instantaneous and average atomic potential energy per atom of $MoS_2$ monolayer sheets with different desulphurization contents, planar dimensions and ratios of length-to-width are recorded to elaborate the desulphurization-induced structural self-assembling behaviors. Self-assembled topological structures of both free-standing and substrate-hosted $MoS_2$ monolayer sheets driven by thermal fluctuations and interatomic interactions strongly depend on the desulphurization contents, planar dimensions and ratios of length-to-width. Interestingly, it is demonstrated that the weak interlayer vdW interactions can also affect the structural morphology of desulphurized $MoS_2$ sheets. Similar to previous reports via experimental observations, $MoS_2$ monolayer sheets with specific conditions including desulphurization and sample dimensions can roll up into nanotube morphology. Besides, some other novel molecular structures, *e.g.*, triangular-tube, diagonal-folding, quadrangle-tube, and amorphous-cluster structures, are also observed, suggesting complex structural morphology phases of desulphurized $MoS_2$ monolayer sheets that are experimental pending. The study not only presents further knowledge of structural properties of 2D $MoS_2$ monolayer sheets, but also potentially provides new opportunities to synthesize TMD with unique morphologies via defect engineering.

## ASSOCIATED CONTENT

### Supporting Information

The Supporting Information is available free of charge at https://xxx.

## AUTHOR INFORMATION

### Corresponding Author

*pinqiang@wust.edu.cn; and jianyang@xmu.edu.cn;

### Author contributions

J. W., and P. C. conceptualized and conceived the project. P. C. performed the molecular dynamics simulations and drew pictures with the important inputs and suggestions from the all co-authors. P. C., and J. W. analyzed and discussed the results. J. W., and P.C. wrote and polished the manuscript.

**Acknowledgments**

This work is financially supported by the National Natural Science Foundation of China (Grant Nos. 11772278 and 11904300), the Jiangxi Provincial Outstanding Young Talents Program (Grant No. 20192BCBL23029), the Fundamental Research Funds for the Central Universities (Xiamen University: Grant Nos. 20720180014 and 20720180018). Y. Yu and Z. Xu from Information and Network Center of Xiamen University for the help with the high-performance computer.

**Competing interests**

The authors declare that they have no competing interests.

**Data and materials availability**

All data needed to evaluate the conclusions in the paper are present in the paper and/or the Supplementary Materials. Additional data related to this paper may be requested from the corresponding authors.

## Figures and Tables

### Figure 1

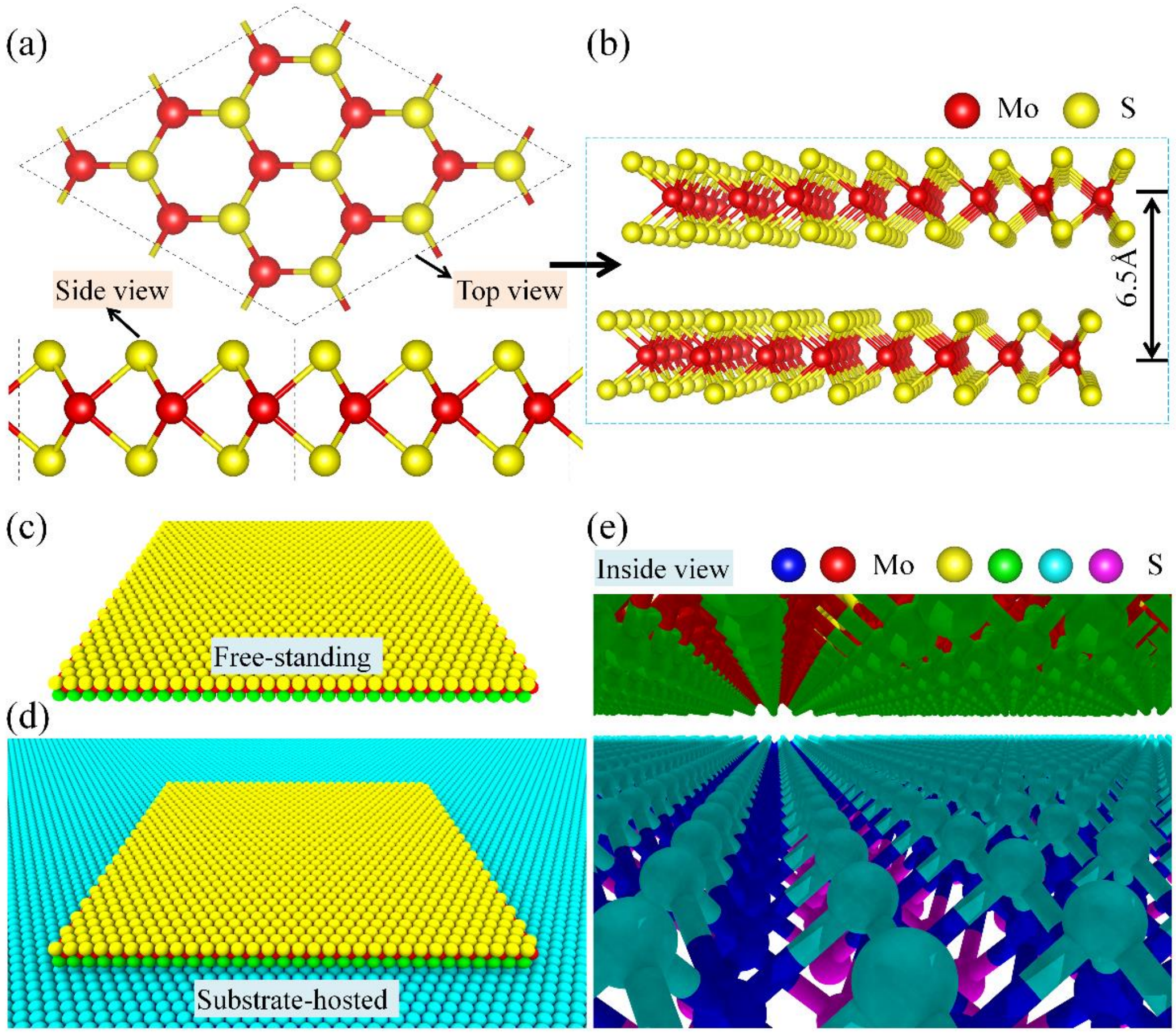

**Figure 1** Atom structures of 2D monocrystalline $MoS_2$ sheets. (a) Top and side views of $MoS_2$ monolayer structures. (b) Representative structure of $MoS_2$ bilayer. (c) A free-standing molecular structure of $MoS_2$ monolayer sheet with a planar dimension of $10 \times 10$ nm$^2$. (d) A substrate-hosted molecular structure $MoS_2$ with a planar dimension of $10 \times 10$ nm$^2$. (e) A zoomed-in structure of substrate-hosted $MoS_2$ sheets at the initial state from the inside view.

**Figure 2**

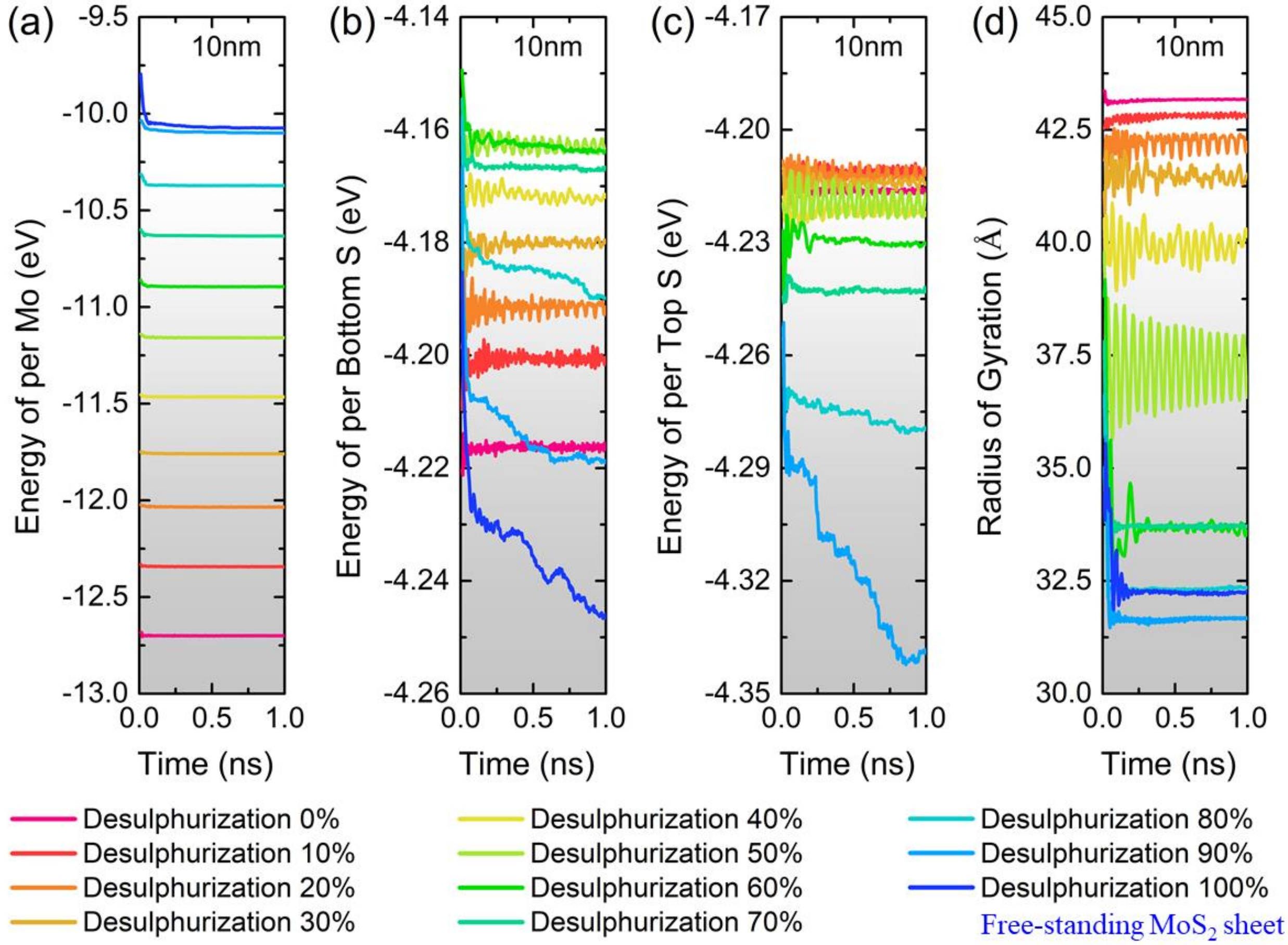

**Figure 2** Energetics in free-standing $MoS_2$ monolayer sheets with a planar dimension of $10 \times 10$ $nm^2$ and desulphurization ranging from 0 to 100%. (a) Relationships between energetics per Mo atom and relaxation time. (b) Relationships between energetics per bottom-S atom and relaxation time. (c) Relationships between energetics per top-S atom and relaxation time. (d) Relationships between radius of gyration and relaxation time.

**Figure 3**

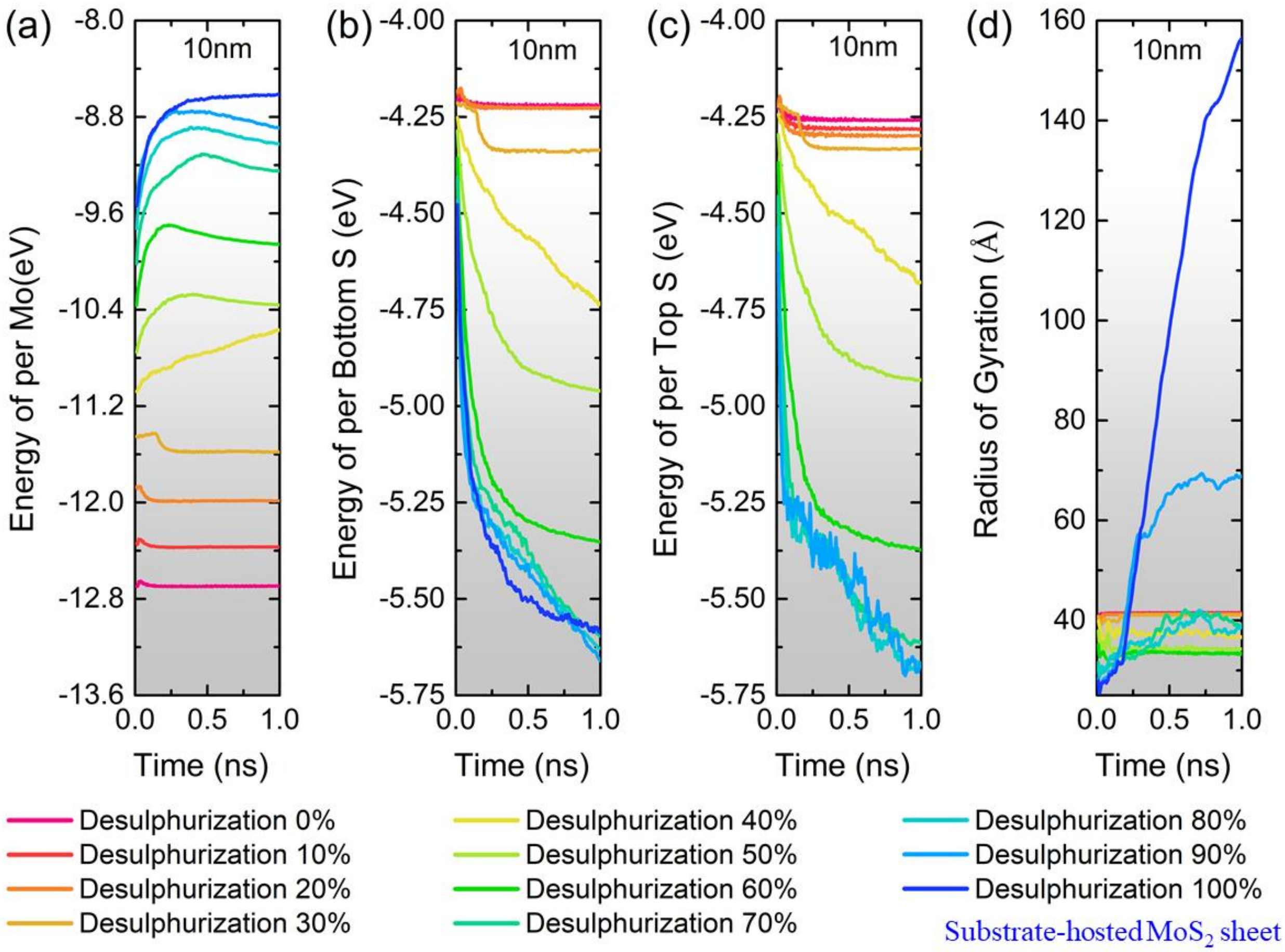

**Figure 3** Energetics in substrate-hosted $MoS_2$ monolayer sheets with a planar dimension of $10 \times 10$ nm$^2$ and desulphurization ranging from 0 to 100%. (a) Relationships between energetics per Mo atom and relaxation time. (b) Relationships between energetics per bottom-S atom and relaxation time. (c) Relationships between energetics per top-S atom and relaxation time. (d) Relationships between radius of gyration and relaxation time.

**Figure 4**

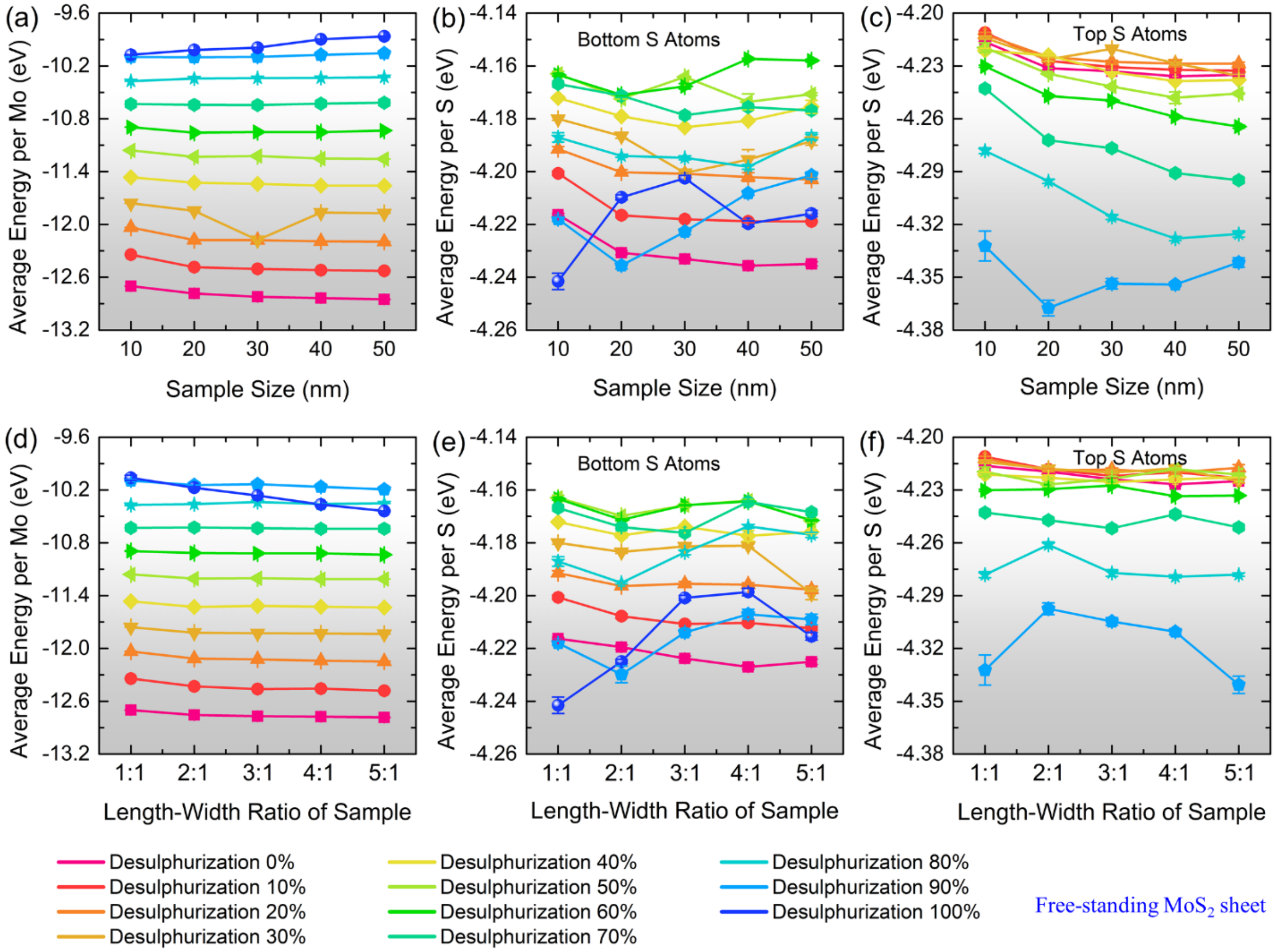

**Figure 4** Average energy-per-atom of free-standing $MoS_2$monolayer sheets. (a)-(c) Average energy-per-atom versus sample sizes. (d)-(f) Average energy-per-atom versus length-width ratios of samples.

## Figure 5

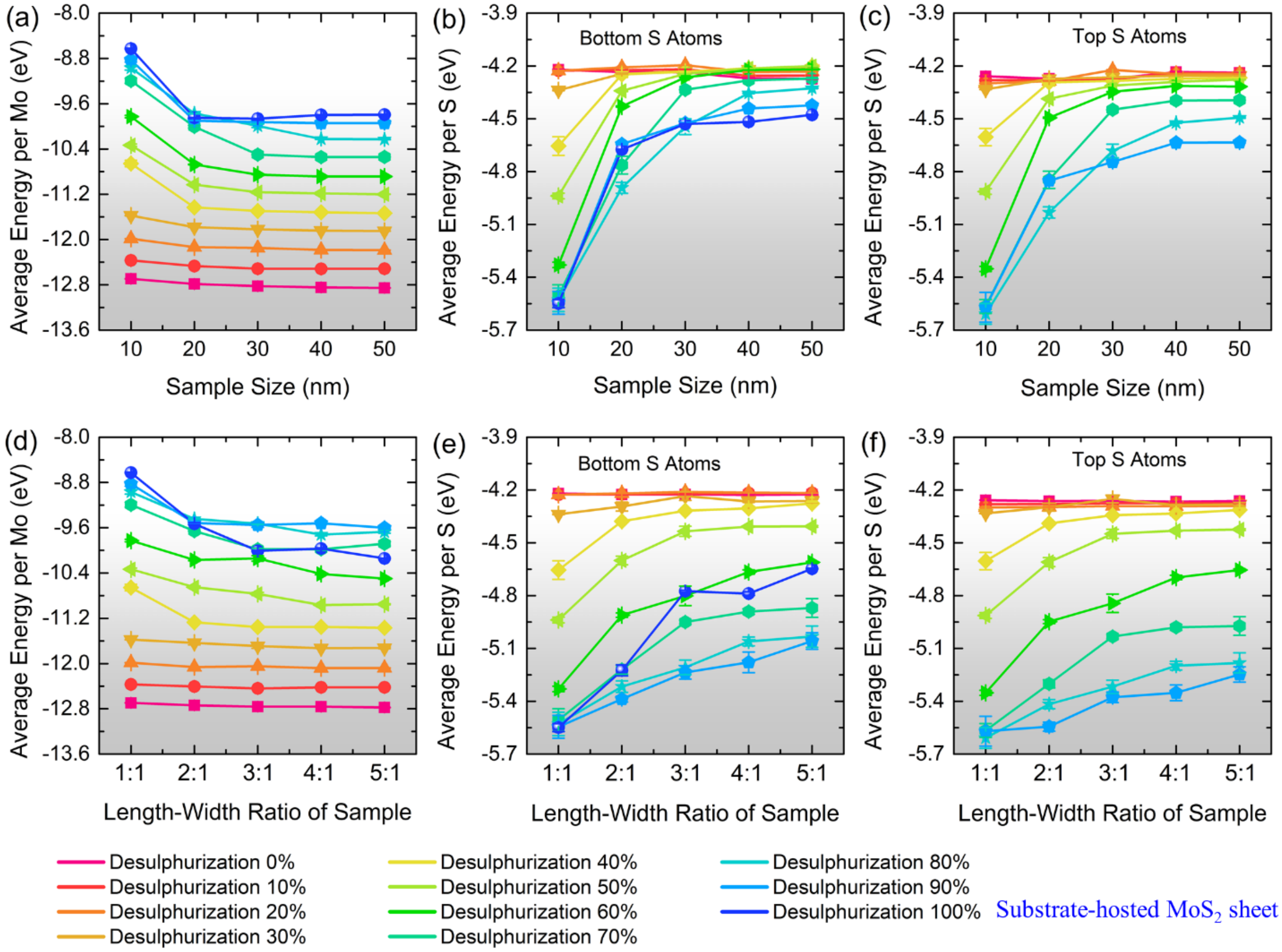

**Figure 5** Average energy-per-atom of substrate-hosted $MoS_2$ monolayer sheets. (a)-(c) Average energy-per-atom versus sample sizes. (d)-(f) Average energy-per-atom versus length-width ratios of sample.

**Figure 6**

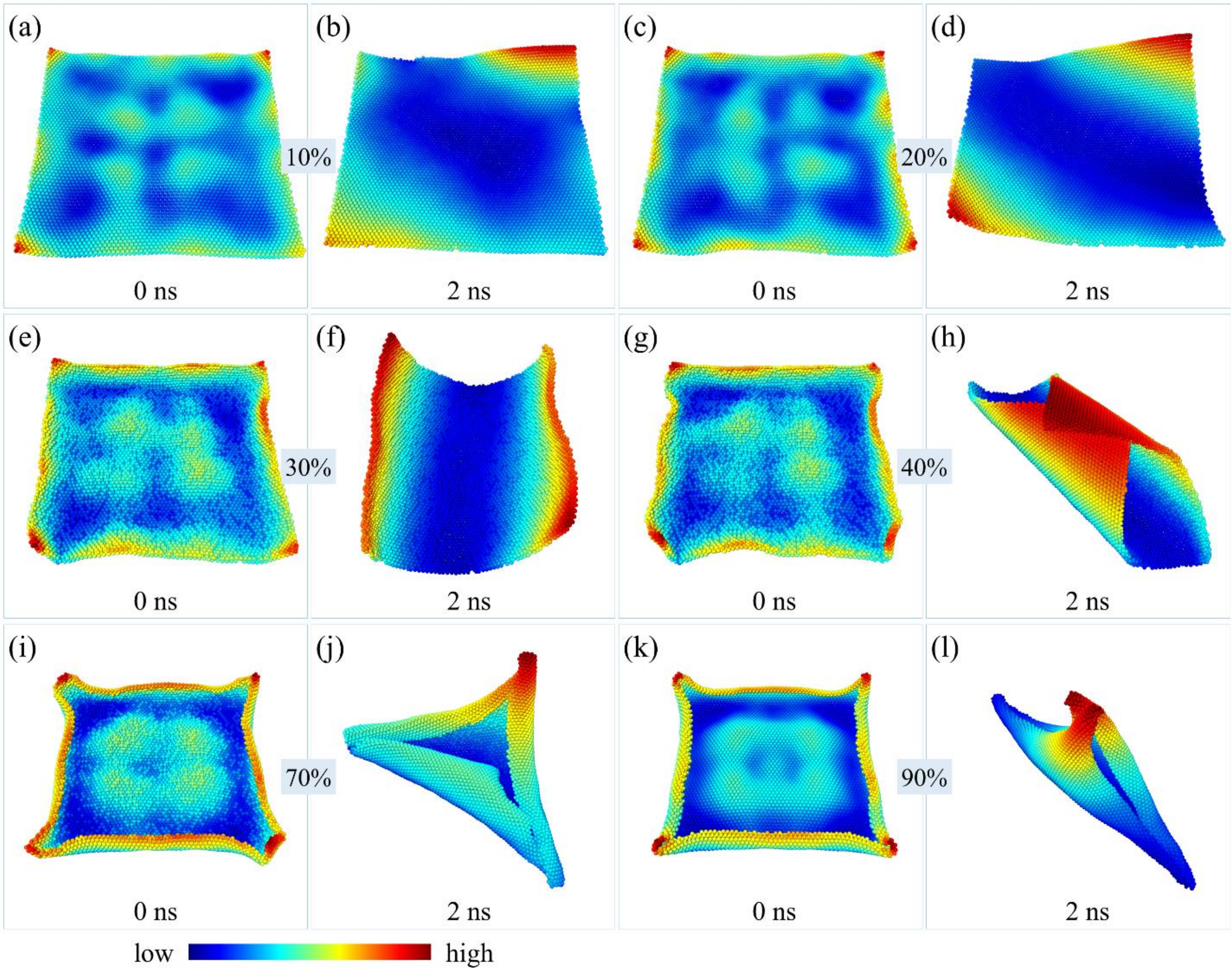

**Figure 6** Structural revolutions of free-standing $MoS_2$ monolayer sheets with a planar dimension of 20 × 20 $nm^2$. Snapshots of free-standing $MoS_2$ monolayer sheets are colored on the basis of the relative displacements in out-of-plane (*z*) direction.

**Figure 7**

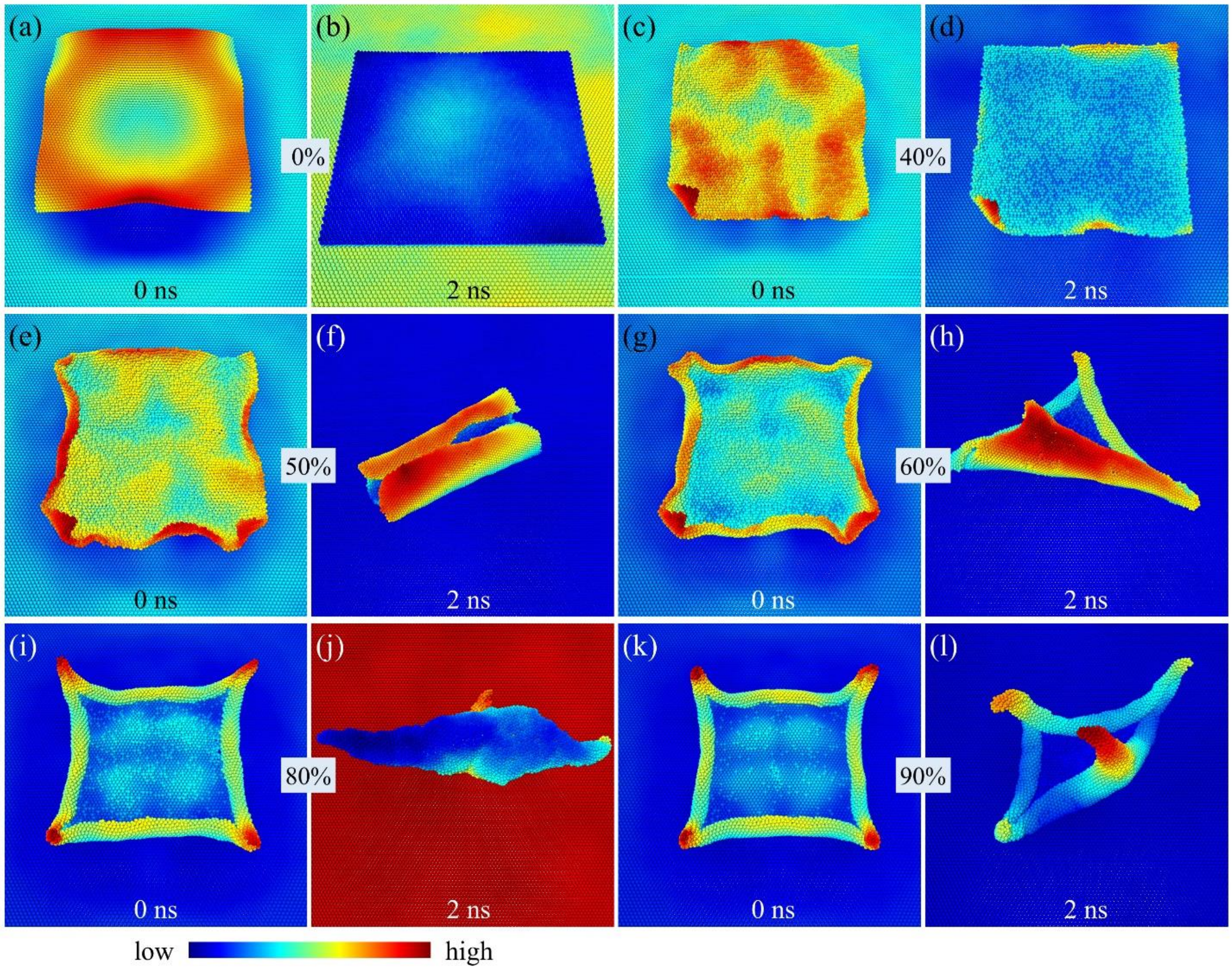

**Figure 7** Structural revolutions of substrate-hosted $MoS_2$ monolayer sheets with a planar dimension of 20 × 20 $nm^2$. Snapshots of substrate-hosted $MoS_2$ monolayer sheets are colored according to the relative displacements in out-of-plane (*z*) direction.

**Figure 8**

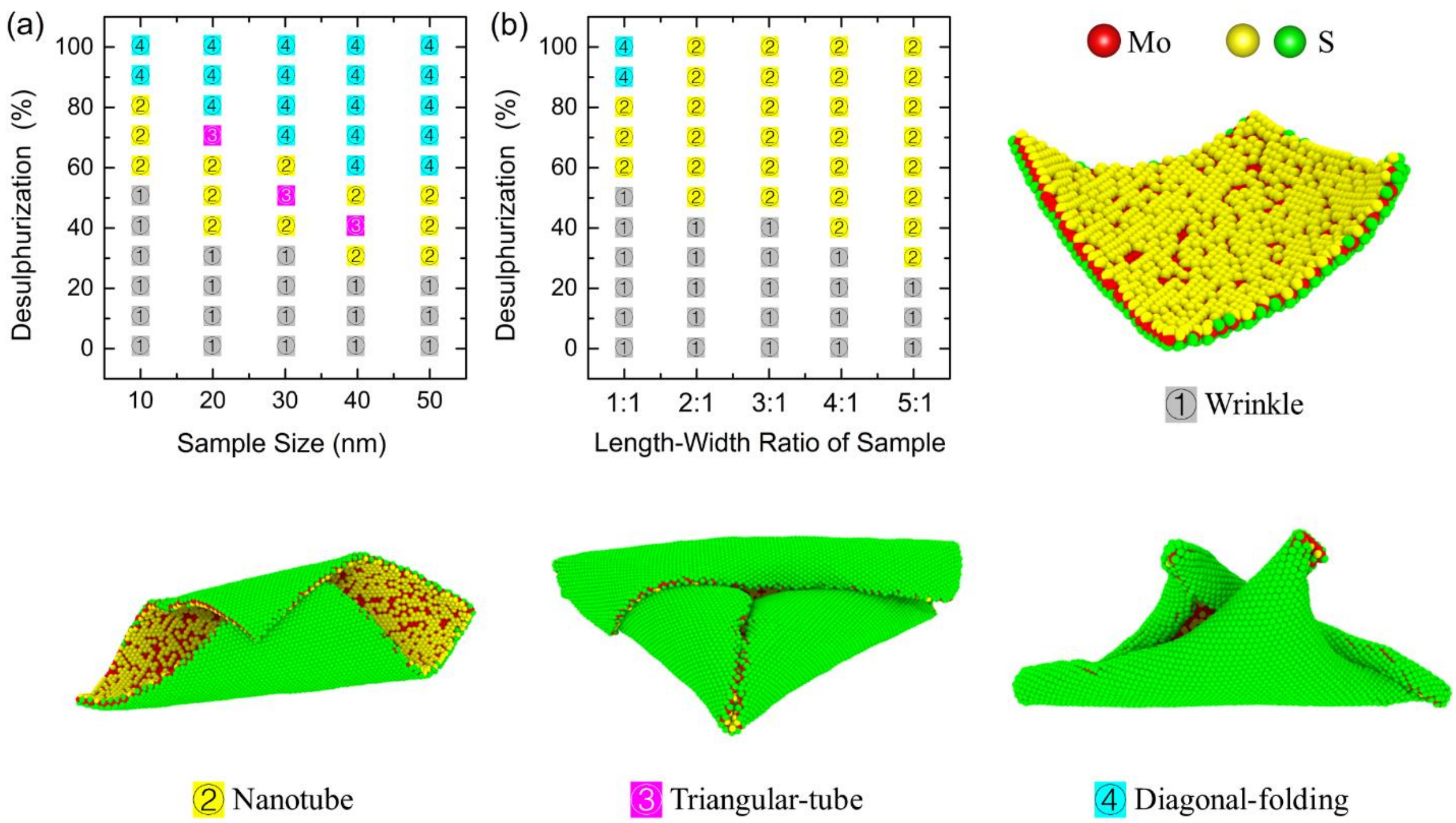

**Figure 8** Molecular morphological diagrams of free-standing $MoS_2$ monolayer sheets. (a) Desulphurization versus sample size diagrams. (b) Desulphurization versus ratios of length-to-width diagrams. ①Relaxed molecular structure with a dimension $10 \times 10$ nm$^2$ and desulphurization of 30%. ② Relaxed molecular structure with a dimension of $20 \times 20$ nm$^2$ and desulphurization of 40%. ③Relaxed molecular structure with a dimension of $30 \times 30$ nm$^2$ and desulphurization of 50%. ④Relaxed molecular structure with a dimension of $20 \times 20$ nm$^2$ and desulphurization of 80%.

**Figure 9**

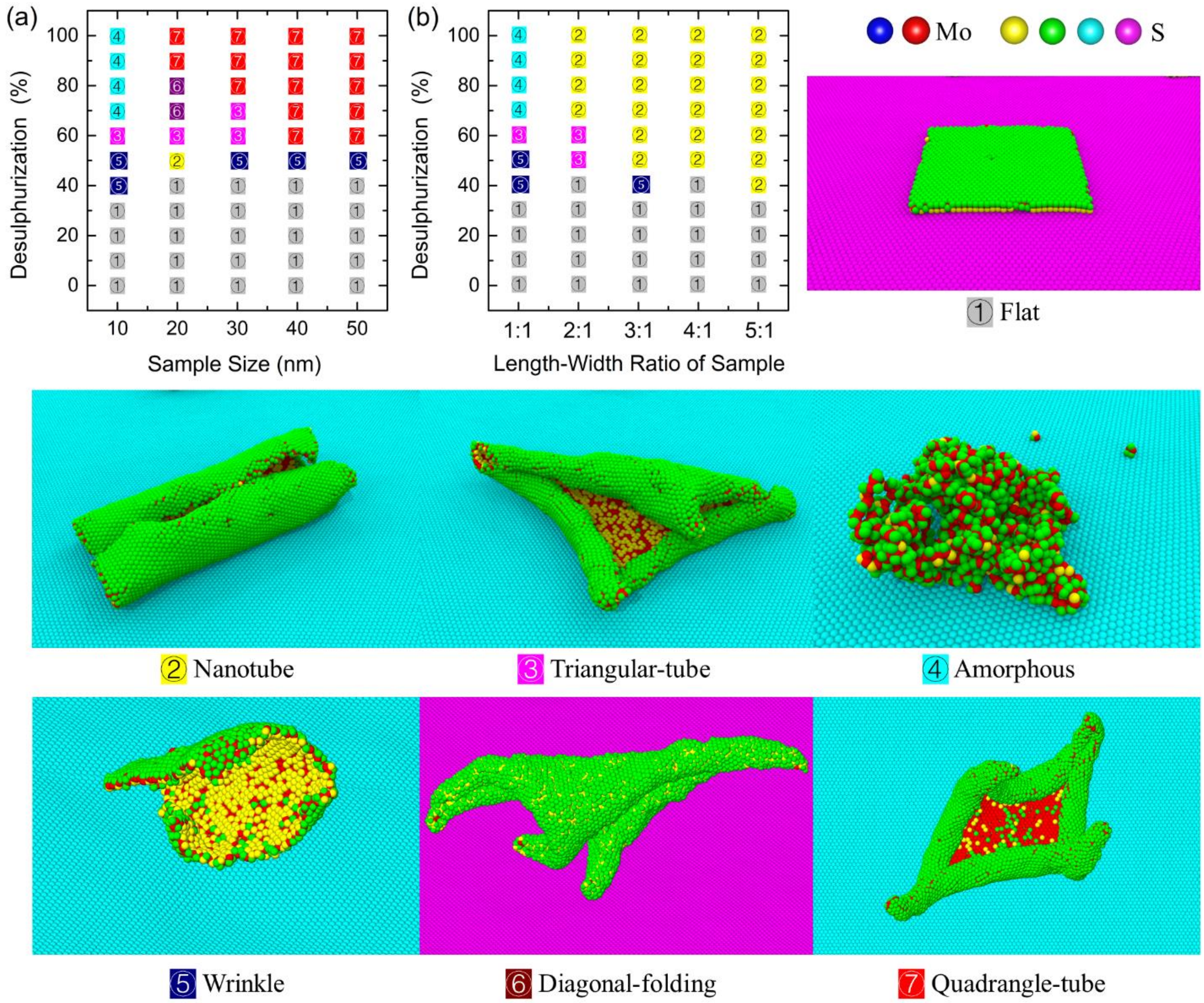

**Figure 9** Molecular morphological diagrams of substrate-hosted $MoS_2$ monolayer sheets. (a) Desulphurization versus sample size diagrams. (b) Desulphurization versus ratios of length-to-width diagrams. ①Relaxed molecular structure with a dimension of 10 × 10 $nm^2$ and desulphurization of 20%. ② Relaxed molecular structure with a dimension of 20 × 20 $nm^2$ and desulphurization of 50%. ③ Relaxed molecular structure with a dimension of 20 × 20 $nm^2$ and desulphurization of 60%. ④ Relaxed molecular structure with a dimension of 10 × 10 $nm^2$ and desulphurization of 90%. ⑤ Relaxed molecular structure with a dimension of 10 × 10 $nm^2$ and desulphurization of 40%. ⑥ Relaxed molecular structure

with a dimension of $20 \times 20$ nm$^2$ and desulphurization of 80%. ⑦ Relaxed molecular structure with a dimension of $20 \times 20$ nm$^2$ and desulphurization of 90%.